\documentclass{iopart}
\usepackage[english]{babel}
\usepackage{iopams}
\pdfoutput=1
\usepackage{setstack}
\usepackage{booktabs}
\usepackage{graphicx}
\usepackage{hyperref}
\usepackage{pdfsync}

\newcommand{\ang}[1]{\mathbb{E}\left [ #1 \right ]}
\newcommand{\abs}[1]{\left| #1 \right|}
\newcommand{\Ac}{\mathcal{A}}

\newcommand{\var}{\mathrm{Var}}
\newcommand{\FY}{F^Y}
\newcommand{\AY}{A^Y}
\newcommand{\aY}{-\kappa_Y}
\newcommand{\bY}{\kappa_Y y_{\infty}}
\newcommand{\cY}{\sigma_Y^2}
\newcommand{\FZ}{F^Z}
\newcommand{\AZ}{A^Z}
\newcommand{\aZ}{-\kappa_Z}
\newcommand{\bZ}{\kappa_Z z_{\infty}}
\newcommand{\cZ}{\sigma_Z^2}
\newcommand{\rYZ}{\rho_{YZ}}
\newcommand{\rXY}{\rho_{XY}}
\newcommand{\rXZ}{\rho_{XZ}}
\newcommand{\ud}{\mathrm{d}}
\newcommand{\ue}{\mathrm{e}}

\begin{document}

\title{Stochastic volatility with heterogeneous time scales}

\author{Danilo Delpini$^{1,2}$ and Giacomo Bormetti$^{3,4}$}
\address{$^1$ Department of Economics and Business (DiSEA), University of Sassari, via Muroni 25, 07100 Sassari, Italy}
\address{$^2$ IMT Institue for Advanced Studies, P.za San Ponziano 6, 55100 Lucca, Italy}
\address{$^3$ Scuola Normale Superiore, Piazza dei Cavalieri 7, 56126 Pisa, Italy}
\address{$^4$ INFN - Sezione di Pavia, via Bassi 6, 27100 Pavia, Italy}
\ead{ddelpini@uniss.it}

\begin{abstract}
  Agents' heterogeneity is recognized as a driver mechanism for the
  persistence of financial volatility.
  We focus on the multiplicity of investment strategies' horizons, we embed
  this concept in a continuous time stochastic volatility framework and prove
  that a parsimonious, two-scale version effectively captures the long memory
  as measured from the real data.
  Since estimating parameters in a stochastic volatility model is challenging, we introduce a robust methodology based on the
  Generalized Method of Moments supported by a heuristic selection of the orthogonal conditions.
  In addition to the volatility clustering, the estimated model also captures
  other relevant stylized facts, emerging as a minimal but realistic and
  complete framework for modelling financial time series.
\end{abstract}
\submitto{Journal of Statistical Mechanics: Theory and Experiment}


\section{Introduction}
In 1963~\cite{Mandelbrot:1963,Mandelbrot:1997} Beno\^it Mandelbrot refers to
the volatility clustering as ``large changes tend to be followed by large
changes, of either sign, and small changes tend to be followed by small
changes''.
Since then this effect has remained one of the most intriguing properties
exhibited by financial time series.
In the early Nineties the long memory property of absolute stock market returns
was independently investigated  by~\cite{Dacorogna_etal:1993}
and~\cite{Ding1993}.
In the former work, after amending absolute price changes from the
heteroscedasticity due to seasonal effects, the authors find a persistent
positive autocorrelation declining hyperbolically with the time lag.
In the latter, analysing the daily closing prices of Standard\&Poor 500 index for the time
span January 3 1928 - August 30 1991, Ding and collaborators study the
power correlation of absolute returns $\abs{r_t}^d$ for positive $d$, finding a
strong persistence especially for $d$ close to one.

The slow decay of the volatility can be ascribed to two rather different mechanisms.
Agent Based Models provide a first explanatory framework, where macroscopic
evidences are explained in terms of microscopic interactions among market
participants.
As clarified in the seminal papers~\cite{Lux_Marchesi:1999,Lux_Marchesi:2000}
the alternation of the economic agents between chartist and fundamentalist
regime can be identified as the source of the observed volatility clustering,
an empirical signature of persistence.
The same mechanism leading to the previous regime switching is further
investigated in~\cite{Alfi_etal:2009a,Alfi_etal:2009b}, where the minimal
assumptions required for an agent based model to capture the empirical stylized
facts  are identified.
In a different approach~\cite{Muller_etal:1994} persistence is induced by the
coexistence of agents differing in their perceptions of the market, risk
profiles, institutional constraints, degree of information, prior beliefs, and
other characteristics such as geographical locations.
In~\cite{Muller_etal:1993} the role of heterogeneous time horizons for the
investment strategies is specifically addressed.
In~\cite{Corsi:2009} the daily, weekly and monthly time scales are isolated as
the relevant ones, while the first direct evidence of these three scales as
well as an attempt to capture them with an ARCH model is provided
in~\cite{Lynch:2003}.
As a major achievement of the latter work we see that a small subset of time
scales succeeds in capturing the long run behaviour of the squared return
correlation.
Interestingly, those horizons reflect typical time scales of the human
activity, which noticeably follow a pseudo-geometric
progression~\cite{Bouchaud_QF:2001}.
Generalizing the concept of a finite mixture of time scales to a continuum of
agents, an attractive intuition is that the integrated effect of exponential
heterogeneous strategies may lead to persistence.
On a formal basis, this amounts to expressing the correlation function as 
\begin{equation*}
  C(\tau)=\int_0^{1/\tau_\mathrm{min}} \exp\left(-\tau/\tau_\mathrm{agent}\right)p(1/\tau_\mathrm{agent})\ud (1/\tau_\mathrm{agent})\,,
\end{equation*}
which at the leading order for $\tau\rightarrow +\infty$ is determined by the
behaviour of the density $p(1/\tau_\mathrm{agent})$ around the origin.
Indeed, by virtue of Watson's Lemma, we obtain $C(\tau)\sim 1/\tau^{1+\alpha}$
provided that $p(1/\tau_\mathrm{agent})\sim \tau_\mathrm{agent}^{-\alpha}$ with
$\alpha> -1$.

As far as the distributional properties of the volatility proxies are concerned,
in~\cite{Micciche:2002} the inverse gamma distribution is identified as
an effective approximation for both the low and high volatility regimes.
The simplest model reproducing this distribution as a result of a volatility
feedback effect corresponds to an ARCH-like equation which, in the continuous
time limit, reads as a Langevin equation
\begin{equation*}
  \frac{\ud \sigma}{\ud t}=-\kappa(\sigma-\sigma_\infty)+\eta\ \sigma\ \zeta(t)\,,
\end{equation*}
with $\kappa$, $\sigma_\infty$, $\eta$ positive constants.
For this specific case, the stationary distribution of the volatility has the
form of an inverse gamma
\begin{equation*}
  \frac{\lambda^\nu}{\Gamma(\nu)}\frac{\ue^{-\lambda/\sigma}}{\sigma^{1+\nu}}
\end{equation*}
with $\nu=1+2\kappa/\eta^2$ and $\lambda=2\kappa\sigma_\infty/\eta^2$.
In the following Section we propose an approach inspired by this evidence about
the volatility, as well as by the idea of a mixture of heterogeneous investment
horizons, in a spirit similar to the Heston multi-factor
model~\cite{Bates:2000}.

The remainder of the paper is organized as follows: in
Section~\ref{sec:nonlinear} we derive analytical expressions for the leverage
and volatility autocorrelation, while in Section~\ref{sec:GMM} we detail a
calibration procedure which is inspired by the Generalized Method of Moments.
We conclude in Section~\ref{sec:conclusions}.
The analytical derivations are postponed in the three Appendices at the end of
the paper.

\section{The model}\label{sec:themodel}
A quite general expression for the asset price at time $t$, reminiscent of the
Geometric Brownian motion paradigm, is given by
\begin{equation*}
	S_t = S_0 \,\exp\left(\mu \,t + X_t\right)
\end{equation*}
where $X_t$ is the stochastic centred log-return and $\mu$ a constant drift
coefficient.
In~\cite{PhysRevE.83.04111} we assume that the time evolution of $X_t$ can be modelled in terms of
the stochastic differential equation (SDE)
\begin{equation}
	\ud X_t = \sigma_t\,\ud W_{t}^X\,,
	\label{eq:dXeq}
\end{equation}
where $\sigma_t$ is the instantaneous volatility of the price and $\ud W_{t}^X$
the increment of a standard Wiener process.
Since $X_0$ is equal to zero, we
also have $\ang{X_t}=0$ and $\ang{\ln{S_t}-\ln{S_0}}=\mu t$ for any $t$.
A common choice accounting for the stochastic behaviour of the volatility, as
measured by suitable proxies,  is $\sigma_t=\sigma(Y_t)$ as a function of an
unobserved driving process $Y_t$.
General financial considerations regarding the mean-reverting behaviour of the
volatility process lead to a second SDE of the form
\begin{equation}
  \ud Y_t = \aY(Y_t -y_\infty)\,\ud t+\sqrt{\Sigma(Y_t)}\,\ud W_{t}^Y\,,
  \label{eq:dYeq}
\end{equation}
with $\kappa_Y=1/\tau_Y>0$, and $y_\infty>0$.
In~\cite{PhysRevE.83.04111} $\Sigma(Y_t)$ is equal to $\sigma_Y^2Y_t^2$ with
$\sigma_Y>0$, from which it follows that $\sigma_t$ is proportional to $Y_t$. This choice leads to an inverse gamma
stationary distribution with shape and scale parameters $\nu=1+2\kappa_Y/\cY$
and $\lambda=2\bY/\cY$, respectively; in light of the considerations presented
in the Introduction, this is dictated by the will of recovering the most effective
statistical description of the volatility distribution.
Different choices for $\Sigma$ have been suggested in the literature and among the most popular ones 
it is worth mentioning the Heston~\cite{Heston1993} and Stein-Stein~\cite{RevFinancStudies.4.727} models.
For a complete overview of continuous time models as well as widely employed discrete time approaches like ARCH, GARCH and their generalizations, 
we suggest the handbook about financial time series~\cite{Andersen_etal:2009}.

Following the spirit of the Introduction, in this paper we extend the model
given by~(1) and~(2) allowing the instantaneous volatility $\sigma_t$ to depend
on multiple stochastic unobserved factors.
Before stating explicitly our model's equations, it is worth noting that such a
generalization is inspired by the multi-factor stochastic volatility model
introduced by Bates in~\cite{Bates:2000} as a possible description of the S\&P
500 futures price, and later revisited in~\cite{Christoffersen_etal} as a model
for the dynamics of the volatility smirk in the option pricing context.
In the generalized form used in~\cite{Corsi:2012}, the multi-factor model with
jumps reads  
\begin{eqnarray}
  \ud X_t & = & \sum_{i=1}^N \sqrt{Y_t^i} \,\ud W^{i}_t + \ud J_t^{X} \nonumber\\
  \ud Y_t^{i} & = & -\kappa_i \left( Y_t^{i} - y_\infty^i \right) \, \ud t + \eta_i \,\sqrt{Y_t^i} \,\ud W^{i+N}_t + \ud J_t^i\,, \quad i=1,\ldots,N \nonumber \\
  &&
  \label{eq:multifactor}
\end{eqnarray}
where $W_t^1,\ldots,W_t^{2N}$ is a multivariate possibly correlated Brownian motion and $\{J^X,J^1,\ldots,J^N\}$ is a multivariate possibly 
correlated Poisson process with constant intensities.
In principle, each factor may be linked to the sensitivity of the economic
agents to different investment horizons, and in light of this heterogeneity the
modelling could reflect $N$ volatility components.
Our starting point is a special case of the dynamics~(3): we take $N=2$,
discard the jump contribution and assume perfect correlation between the Wiener
processes of the two factors, $\mathrm{corr(W_t^2, W_t^2)}=1$.
However, at variance with equation~(3) and as a major contribution of our
paper, we consider inverse gamma driving factors, each one being described by
the same mean-reverting dynamics provided in (2) with $\Sigma$ proportional to
the squared process.
Ultimately the model we are going to analyse reduces to
\begin{eqnarray}
  \ud X_t & = & Y_t \,\ud W^{X}_t + Z_t \,\ud W^{X}_t\nonumber\\
  \ud Y_t & = & \aY \left( Y_t - y_\infty \right) \, \ud t + \sigma_Y \,Y_t \,\ud W^Y_t \nonumber\\
  \ud Z_t & = & \aZ \left( Z_t - z_\infty \right) \, \ud t + \sigma_Z \,Z_t \,\ud W^Z_t \,,
  \label{eq:model}
\end{eqnarray}
where we impose the initial time conditions $X_{t=0}=X_0=0$,
$Y_{t=t_0}=y_0>0$ and $Z_{t=t_0}=z_0>0$, with $\kappa_Y=1/\tau_Y>0$, and
$\kappa_Z=1/\tau_Z>0$.
We also indicate $\nu_Y=1+2\kappa_Y/\sigma_Y^2$ and
$\nu_Z=1+2\kappa_Z/\sigma_Z^2$ the tail exponents of the inverse gamma
stationary distributions of $Y_t$ and $Z_t$.
The correlation structure among the three Brownian motions is described by the
following matrix
\begin{displaymath}
  \left(
  \begin{array}{ccc}
    1 & \rXY & \rXZ\\
    \rXY & 1 & \rYZ\\
    \rXZ & \rYZ & 1    
  \end{array}
  \right)\, .
\end{displaymath}

It has to be noted that we assume different starting times for the volatility
factors and the return process, according to what is done
in~\cite{PhysRevE.83.04111}.
Indeed, as the processes $Y_t$ and $Z_t$ are unobserved factors and we
are mainly concerned with their dynamics at the stationary state, we
assume they start at $t_0<0$ in the past and we recover the
stationary limit by letting $t_0\to -\infty$.
On the other hand, $X_t$ represents the observed (detrended) logarithmic
increment of the price for a fixed time lag and, therefore, it seems natural to
take the spot time $t=0$ as a starting time for this lag.

Some considerations are due regarding our choice of the factors specification.
In~\cite{PhysRevE.83.04111} the single factor $Y_t$ corresponds (up to a
constant) to the instantaneous volatility itself.
As such, $Y_t$ has a clear interpretation and its dynamics is chosen
specifically with the intent to accommodate the distributional properties of
the volatility observed in the reality.
Here, in the spirit of the factor model~\eref{eq:multifactor}, the evolution of
log-returns is given in terms of two additive factors; it follows that $Y_t$
and $Z_t$ can not be interpreted, separately, as the return volatility (or
possibly the variance as in ARCH/GARCH models), but as the underlying unobserved factors. 
The choice of full correlation between $W_t^1$ and $W_t^2$ allows us to
introduce formally $\sigma_t = Y_t + Z_t$ and to motivate the inverse gamma
dynamics which drives the factors.
Implicitly, this means that we give up recovering exactly an asymptotic inverse
gamma law for $\sigma_t$ and we give priority to capturing the observed, long
range memory of the squared return correlation.
Nevertheless, we expect the tail asymptotic to be preserved under suitable
assumptions (see discussion at the end of this Section). 
Finally, we observe that generalization with more that two factors is
straightforward, but cumbersome, and would greatly simplify if we assume
$\mathrm{corr}(W_t^1, W_t^2)=0$ in~(\ref{eq:multifactor}).
However, the specific purpose of this work is to show how our minimal
choice is indeed able to capture the very consequences of heterogeneity.

From~\cite{PhysRevE.83.04111} we know that a negative $\rXY$ suffices to
accommodate the observed short range scaling of the return-volatility
correlation; in~\ref{app:B} and in the numerical section we set $\rXZ$ equal to zero to prevent $Z_t$ from impacting
the leverage.  Nonetheless, in what follows we derive  the relation between the factors behaviour and the moments of $X_t$ under the general case 
of non trivial correlations between the Brownian motions.
As we show in \ref{app:A}, the structure of the model~\eref{eq:model} allows to
compute the moments of the probability density function (PDF) of $X_t$ at all
times $t$ recursively.
After cumbersome calculations, and by exploiting It\^o's Lemma to compute
the cross correlations between the two volatility factors, 
it can be verified that the moments of $X$ can be expressed always as a
superposition of exponential functions of $(t-t_0)$
\begin{equation}
  \ang{X^n_t} = \sum_{i,j=0;~i+j\le n}^{n} H^{(n)}_{i,j}(t;y_0, z_0)
	\,\exp{\left(F_{i,j}(t- t_0)\right)} \,,
  \label{eq:XmomsExpand}
\end{equation}
where the constants read $F_{m,n}=\FY_m + \FZ_n +m n \rYZ \,\sqrt{\cY
\cZ}$, with $\FY_m= \aY m + m(m-1) \cY/2$, and $\FZ_n=\aZ n + n(n-1)\cZ/2$.
The coefficients $H^{(n)}_{i,j}$ depend on the time lag $t$; more precisely,
due to the linearity of the ODEs~\eref{eq:corrDiffEq}, they correspond to a
combination of exponential terms weighted by polynomials in $t$.

In the following, we report the explicit expressions of the coefficients
$H^{(n)}_{i,j}(t; y_0, z_0)$ for the case $n=2$ (the constants
$k^{(m,n)}_{i,j}$, which depend on the initial conditions $y_0, z_0$,
are defined recursively in~\ref{app:A})
\begin{eqnarray*}
  H^{(2)}_{0,0} & = & \left[k^{(2,0)}_{0,0}+2k^{(1,1)}_{0,0}+k^{(0,2)}_{0,0}\right] t\,, \\
  H^{(2)}_{1,0} & = & \left[k^{(2,0)}_{1,0}+2k^{(1,1)}_{1,0}\right] \frac{ 1 - \exp{\left(-F_{1,0} t\right)}}{F_{1,0}}\,,\\
  H^{(2)}_{0,1} & = & \left[k^{(2,0)}_{0,1}+2k^{(1,1)}_{0,1}\right] \frac{1-\exp{\left(-F_{0,1} t\right)}}{F_{0,1}}\,,\\
  H^{(2)}_{2,0} & = & k^{(2,0)}_{2,0} \frac{1-\exp{\left(-F_{2,0} t\right)}}{F_{2,0}}\,,\\
  H^{(2)}_{1,1} & = & 2 k^{(1,1)}_{1,1} \frac{1-\exp{\left(-F_{1,1} t\right)}}{F_{1,1}}\,,\\
  H^{(2)}_{0,2} & = & k^{(0,2)}_{0,2} \frac{1-\exp{\left(-F_{0,2} t\right)}}{F_{0,2}}\,.
\end{eqnarray*}
Since $t$ is finite, the coefficients $H^{(n)}_{i,j}$ are finite quantities
themselves, and all the relevant information about the behaviour of
$\ang{X^n_t}$ in the stationary limit of $Y$ and $Z$ is retained by the
$t_0$-exponentials in \Eref{eq:XmomsExpand}.
Given that $F_{0,0}=0$, if all the $F_{i,j}$ for $i,j=0,\ldots,n$ with $i+j\leq
n$ are negative, $\ang{X^n_t}$ is finite in the stationary limit
$t_0\to-\infty$, otherwise it diverges indicating the emergence of fat tails in
the PDF $p_t(x)$ of $X_t$.
In the latter case the tail behaviour would be compatible with an hyperbolic
scaling with a tail exponent smaller than the order of the lowest diverging
moment.

In~\cite{PhysRevE.83.04111} the hyperbolic scaling of $p_t(x)$ is induced by
the power-law tail of the asymptotic (inverse gamma) distribution of
the volatility, and a simple relation exists between the tail exponent of the
latter and the order of the first diverging moment of $p_t(x)$.
In the present case, the asymptotic distribution of $\sigma_t$ is that of the sum
of the two factors $Y$ and $Z$, both inverse gamma distributed with tail
indices $\nu_Y$ and $\nu_Z$ respectively.
In the limit of $Y$ independent of $Z$\footnote{In fact, we make this assumption
when estimating the model from the empirical data in Section~\ref{sec:GMM}.},
the distribution of the sum behaves as a power-law with tail index $\nu_{\sigma}=\min\{\nu_Y, \nu_Z\}=\nu$,
see e.g.~\cite{Bouchaud_book:2003,Wilke:1998}.
Therefore, the same mechanism discussed in~\cite{PhysRevE.83.04111}, which
triggers the divergence of the return moments, applies here asymptotically in
the absolute value $\abs{X}$, and the PDF $p_t(x)$ manifests a decay compatible
with a power-law scaling with tail index determined by the value of
$\nu_{\sigma}$.

The theoretical PDF is also compliant with the more basic properties of the
returns.
In particular, the stationary limit of~\eref{eq:XmomsExpand}, and the
expression of $H^{(2)}_{0,0}$, show that the variance is linear in the
return time lag $t$.
Furthermore, the explicit expressions of the functions $H^{(3)}_{0,0}$ and
$H^{(4)}_{0,0}$, not reported here for the sake of parsimony, would also reveal
that the skewness and kurtosis vanish in the limit of large $t$, according to
the observed Gaussian-like shape of the distribution for large time horizons.

\section{Non linear dependence}\label{sec:nonlinear}
In this Section we discuss the main properties of the return correlation
structure predicted by model~\eref{eq:model}, focusing on the
return-volatility and the squared-return correlation functions.

Model~\eref{eq:model} inherits from the class of stochastic volatility models
the important property of absence of serial correlation, which is verified
empirically with good approximation.
Despite this, financial returns can not just follow a random walk process,
since this would imply independent and identically distributed price
increments.
Any non linear function of the returns would then exhibit zero autocorrelation,
a property that simply does not hold in practice.
Empirical evidences of this violation are the \textit{leverage effect} and the
\textit{volatility clustering}.
The former refers to the negative correlation between past returns and the
future instantaneous volatility, measuring the tendency of the market
volatility to increase after a price
downfall~\cite{PRL.87.228701,Bouchaud_book:2003,Perello:2004}; volatility
clustering is usually expressed in terms of the persistent correlation between
squared returns or logarithm of absolute returns, implying that large variations
are more likely to be followed by large than small
ones~\cite{Dacorogna_etal:1993,Guillaume_etal:1997,Cont_Potters_Bouchaud:1997,Liu_etal:1997,Muzy_Delour_Bacry:2000}.
For a survey of contributions on the same topics from the econometric
community, we refer the interested readers to the reference list
in~\cite{Bouchaud_QF:2001,Cont_QF:2001}.
Our model deals explicitly with these non linear correlation functions; their
expressions, whose derivation we postpone to \ref{app:A} - \ref{app:C}, provide a
valuable analytical characterization of model~(\ref{eq:model}).
This information can be exploited for the calibration of the model from
empirical data, but it is also crucial to grasp the relevant information about
the process time scaling, as we discuss in the rest of this Section.

\subsection{Leverage effect}\label{sec:leverage}
The leverage, a measure of the correlation between returns and volatility, is
usually defined as $\mathcal{L}(\tau;t) = \ang{\ud X_t \,\ud
X_{t+\tau}^2}/\ang{\ud X_t^2}^2$.
Empirically and for arbitrary $t$, $\mathcal{L}(\tau; t)$ has been found to be
negative and exponentially decaying for positive $\tau$ and approximately zero
otherwise: a correlation exists between past returns and the volatility in the
future and not \emph{vice versa}.

For our case, a finite time, exact expression is derived in \ref{app:B};
\Eref{eq:Lev} reveals that the leverage function is characterized by the
superposition of three exponential functions, with different characteristic
times $\tau_\mathcal{L}$, $\tau_\mathcal{\sim L}$, $\tau_{<}$, which are
ordered according to the following hierarchy:
\begin{eqnarray*}
  \tau_\mathcal{L} & = &\frac{2}{2-\tau_Y\sigma_Y^2}\tau_Y=\frac{\nu_Y-1}{\nu_Y-2}\tau_Y\,;\\
  \tau_\mathcal{\sim L} & = &\left(\frac{1}{\tau_Y}-\frac{\sigma_Y^2}{2}+\frac{1}{\tau_Z}\right)^{-1}
  =\frac{\tau_Z}{\tau_Z+\tau_\mathcal{L}}\tau_\mathcal{L}<\tau_\mathcal{L}\,;\\
  \tau_{<}  & = &\left[2\left(\frac{1}{\tau_Y}-\frac{\sigma_Y^2}{2}\right)-\frac{\sigma_Y^2}{2}\right]^{-1}
  =\frac{2}{4-\tau_\mathcal{L}\sigma_Y^2}\tau_\mathcal{L}=\frac{\nu_Y-1}{2\nu_Y-3}\tau_\mathcal{L}\,.
\end{eqnarray*}
If $\nu_Y\rightarrow 3^+$, $\tau_\mathcal{L}$ converges to $2\tau_Y$, while for
$\nu_Y\rightarrow +\infty$ we have that $\tau_\mathcal{L}$ goes to $\tau_Y$.
The time scale $\tau_\mathcal{\sim L}$ is strictly smaller than
$\tau_\mathcal{L}$; however, we are implicitly assuming that the characteristic
time of the $Z$ process accounts for the volatility persistence, that is
$\tau_Z \gg \tau_{\mathcal{L}}$, implying that $\tau_\mathcal{\sim L}$ is
expected to be only slightly smaller than the leverage scale.
Ultimately, if $\nu_Y\rightarrow 3^+$, $\tau_<$ converges to
$2\tau_\mathcal{L}/3$, while under the Gaussian limit we have that $\tau_<$
converges to $\tau_\mathcal{L}/2$.
In fact, the three leverage scales are constrained in a narrow range, which
empirically has been found to be of order ten days for indexes, or even larger for
single stocks~\cite{Bouchaud_book:2003}.

\subsection{Autocorrelation function of squared increments}\label{ssec:autocorr_vola}
The volatility clustering is commonly measured by the quantity $\ang{\ud X_t^2
\,\ud X_{t+\tau}^2}$ and the volatility autocorrelation can be estimated in
terms of the following normalized quantity
\begin{equation}
  \Ac(\tau;t) = \frac{\ang{dX_t^2 \,dX_{t+\tau}^2}-\ang{dX_t^2}
  \ang{dX_{t+\tau}^2}}{\sqrt{\var[dX_t^2]\, \var[dX_{t+\tau}^2]}}\,.
  \label{eq:AcorrDef}
\end{equation}
To complete our analytical characterization of the two-factor
model~\eref{eq:model}, in~\ref{app:C} we derive the explicit expression
of~\eref{eq:AcorrDef}.
At variance with the one previously found in~\cite{PhysRevE.83.04111} and which
is unable to capture the persistence of the volatility, the
expression~\eref{eq:ACFnum3} features five different exponential scales
$\tau^{(i=1,\ldots,5)}_{\Ac}$.
Similarly to the leverage, the characteristic times are organized in a
hierarchy as follows
\begin{eqnarray*}
  \tau^{(1)}_{\Ac} & = & -\frac{1}{F_2^Z} = \frac{\tau_\mathcal{L}}{\tau_Y}\tau_Z = \tau_{>Z} > \tau_Z\,,\\
  \tau^{(2)}_{\Ac} & = & -\frac{1}{F_1^Z} = \tau_Z\,,\\
  \tau^{(3)}_{\Ac} & = & -\frac{1}{F_1^Y} = \tau_Y<\tau_\mathcal{L}\,,\\
  \tau^{(4)}_{\Ac} & = & -\frac{1}{(F_1^Y+F_1^Z)} = \frac{\tau_Z}{\tau_Z+\tau_Y}\tau_Y<\tau_Y\,,\\
  \tau^{(5)}_{\Ac} & = & -\frac{1}{F_2^Y} = \frac{\tau_\mathcal{L}}{2}\,.\\
\end{eqnarray*}
For $\nu_Y$ varying in $(4,+\infty)$, $\tau_Y$ is inferiorly bounded by
$2\tau_\mathcal{L}/3$, while the upper bound is given by $\tau_\mathcal{L}$.
Therefore $\tau_{>Z}$ ranges between $\tau_Z$ and $3\tau_Z/2$, and we can
conclude that the previous five scales indeed cluster into two groups, a
long-range and a short-range one: the first set is $\{\tau_Z, \tau_{>Z}\}$,
whose typical scale is given by $\tau_Z$, while the second one contains the
three remaining scales, superiorly bounded by $\tau_\mathcal{L}$ and of order
$\tau_Y$.
Ultimately, we can appreciate the very reason why model~\eref{eq:model} has been enriched by a second factor process $Z_t$ \textit{w.r.t.} the one
proposed and discussed in~\cite{PhysRevE.83.04111}.
Through the coupling provided by $\rXY\neq 0$, $Y_t$ is entirely responsible
for the emergence of the leverage; conversely, the Brownian motion driving
$Z_t$ is decoupled from $W^X_t$, can not interfere with the leverage, and can not
constrain its hierarchy of time scales, as it happens
in~\cite{PhysRevE.83.04111}.
In model~\eref{eq:model} $Z_t$ provides the degree of freedom required to
capture the persistence of volatility.
It is not difficult to imagine that extra volatility factors would induce a new
plethora of time scales.
However, even though the analytical tractability would be preserved, this would
come at the cost of an overwhelming burden of messy calculations.

\section{Calibration via Generalized Method of Moments}\label{sec:GMM}
In this Section, we propose and discuss an application of the Generalized
Method of Moments to the estimation of the model's parameters from en
empirical time series of price increments.

\begin{figure}[h]
  \caption{\label{fig:ACorr}Empirical volatility autocorrelation function of
  the daily returns of the S\&P500 index 1970-2010 (data points), and
  analytical descriptions: bold and dotted lines, new expressions with GMM estimates for different values of $K'$; dashed
  line, formula and values of parameters as in~\cite{PhysRevE.83.04111}.}
  \begin{center}
    \includegraphics[width=.79\textwidth]{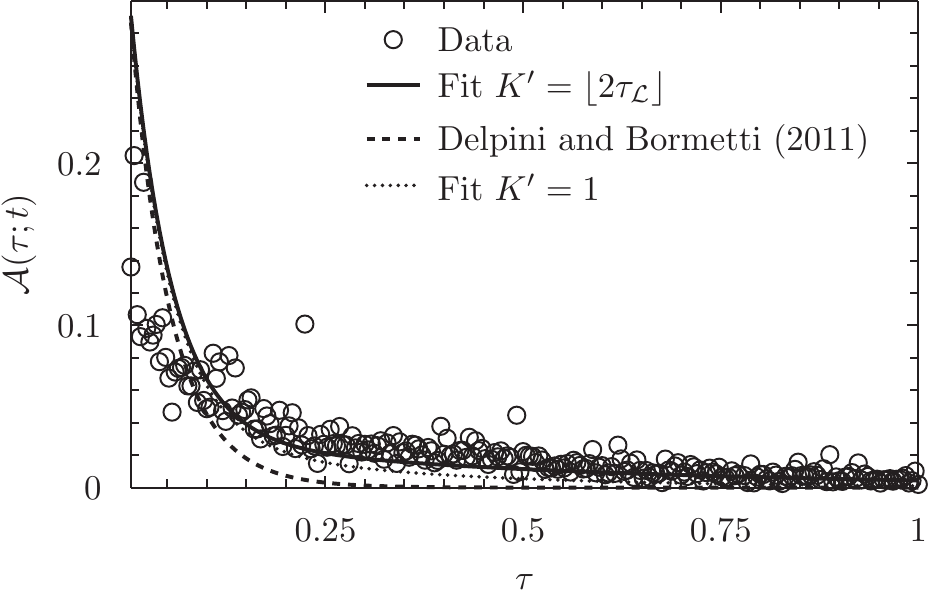}
  \end{center}
\end{figure}

The stochastic model~\eref{eq:model} is characterized by twelve free
parameters, $\tau_Y$, $y_\infty$, $y_0$, $\sigma^2_Y$, $\tau_Z$, $z_\infty$,
$z_0$, $\sigma^2_Z$, $\rXY$, $\rXZ$, $\rYZ$, and $\mu$.
Estimating parameters in a stochastic volatility model is a challenging task.
This is primarily due to the latency of the volatility state variable.
Indeed, in different approaches to volatility modelling, like ARCH and GARCH
models, the likelihood function is readily available.
This problem has inspired many scholars, and there is a specialized literature
on computationally intensive methods mimicking likelihood-based inference.
In general, these belong to the class of non linear filtering methods, and
among possible approaches we mention Kalman filters, Particle filters, and
Monte Carlo Markov Chain approaches.
For more techniques and further discussion we refer the reader to the
handbook~\cite{Andersen_etal:2009}.
Here we take the opportunity to quote the interesting proposal discussed
in~\cite{Javaheri:2005} and rooted on the spectral approach to nonlinear
filtering.
A relatively simpler approach to estimation, which does not rely on any ad hoc
approximation of the density of returns, is based on the computable moments of
the model.
For continuous-time stochastic volatility models, it is generally very hard to
derive closed form solutions for the return moments, but this is not the case
for the model under consideration.
For this reason, we follow a methodology inspired by the Generalized Method of Moments (GMM).
An introduction to the GMM, based on Hansen's formulation of the estimation
problem~\cite{Hansen1982}, is provided by~\cite{Hamilton1994} in Chapter 14.
Given $T$ observations $\{\mathbf{W}_t\}$ for $t=1,\ldots,T$, each one being an
$h$ dimensional vector, and a vector $\boldsymbol{\theta}\in\mathbb{R}^k$ of
unknown parameters, in order to apply GMM there should be a function
$\mathbf{h}(\boldsymbol{\theta},\mathbf{W}_t):\mathbb{R}^k\times\mathbb{R}^h\rightarrow\mathbb{R}^r$
characterized by the property that 
\begin{equation}
  \ang{\mathbf{h}(\boldsymbol{\theta},\mathbf{W}_t)} = \mathbf{0}\,.
\end{equation}
These $r$ equalities are usually described as orthogonality conditions.
The basic idea of GMM is to replace these conditions with sample averages and
to solve the following optimization problem
\begin{equation}
  \hat{\boldsymbol{\theta}} =\underset{\boldsymbol{\theta\in\Theta}}{\mathrm{argmin}} \left(\frac{1}{T}\sum_{t=1}^T\mathbf{h}(\boldsymbol{\theta},\mathbf{W}_t)\right)^{\mathrm{t}}
  \hat{\boldsymbol{\Omega}}_T^{-1}\left(\frac{1}{T}\sum_{t=1}^T\mathbf{h}(\boldsymbol{\theta},\mathbf{W}_t)\right)\,,
\end{equation}
where $\hat{\boldsymbol{\Omega}}_T$ is a positive-definite weighting matrix
depending on the available data set and on the value of $\boldsymbol{\theta}$
itself.
The practical procedure is the one which follows. An initial estimate
$\hat{\boldsymbol{\theta}}^{(0)}$ is obtained by minimizing the previous
quantity with an arbitrary choice of $\hat{\boldsymbol{\Omega}}_T$, e.g.
$\hat{\boldsymbol{\Omega}}_T=\boldsymbol{\mathbb{I}}_{r\times r}$.
Supposing that $\mathbf{h}(\boldsymbol{\theta},\mathbf{W}_t)$ is serially
uncorrelated, the estimate $\hat{\boldsymbol{\theta}}^{(0)}$ is then used in
\begin{equation*}
  \hat{\boldsymbol{\Omega}}_T = \frac{1}{T}\sum_{t=1}^T\mathbf{h}(\hat{\boldsymbol{\theta}}^{(0)},\mathbf{W}_t)\mathbf{h}^{\mathrm{t}}(\hat{\boldsymbol{\theta}}^{(0)},\mathbf{W}_t)
\end{equation*}
to arrive to a new GMM estimate $\hat{\boldsymbol{\theta}}^{(1)}$.
This process can be iterated until an arbitrary stopping criterion is
invoked~\footnote{When the process
$\mathbf{h}(\boldsymbol{\theta},\mathbf{W}_t)$ for $t=1,\ldots,T$ is serially
correlated, the Newey-West estimate for $\hat{\boldsymbol{\Omega}}_T$ can be
used, please refer to equation 14.1.19 in~\cite{Hamilton1994} for further
details.}.
If $\bar{\boldsymbol{\theta}}$ denotes the true value of $\boldsymbol{\theta}$,
the theory behind the GMM states that $\hat{\boldsymbol{\theta}}^{(1)}$ is
approximately distributed as
$\mathrm{Normal}(\bar{\boldsymbol{\theta}},\hat{\boldsymbol{V}}_T/T)$ with
\begin{equation*}
  \hat{\boldsymbol{V}}_T=\left\{
  \underset{(k\times r)}{\left.\frac{\partial}{\partial\boldsymbol{\theta}}\left(\frac{1}{T}
  \sum_{t=1}^T\mathbf{h}^\mathrm{t}(\boldsymbol{\theta},\mathbf{W}_t)\right)\right|_{\boldsymbol{\theta}=\hat{\boldsymbol{\theta}}^{(1)}}}
  \hat{\boldsymbol{\Omega}}_T^{-1}
  \underset{(r\times k)}{\left.\frac{\partial}{\partial\boldsymbol{\theta}^\mathrm{t}}\left(\frac{1}{T}
  \sum_{t=1}^T\mathbf{h}(\boldsymbol{\theta},\mathbf{W}_t)\right)\right|_{\boldsymbol{\theta}=\hat{\boldsymbol{\theta}}^{(1)}}}
  \right\}^{-1}\,.
\end{equation*}
For the case under consideration we have
$\boldsymbol{\theta}^\mathrm{t}=\left(\mu,\tau_Y,y_\infty,y_0,\sigma^2_Y,\tau_Z,z_\infty,z_0,\sigma^2_Z,\rXY,\rXZ,\rYZ\right)$,
while the orthogonality conditions can be obtained computing the lowest order
moments of returns, the leverage correlation, and the squared return
autocorrelation
$$
  \ang{\mathbf{h}(\boldsymbol{\theta},\boldsymbol{W}_{t})}=\ang{
  \begin{array}{c}
    \Delta X_t\\
    \\
    \abs{\Delta X_t} - \sqrt{\frac{2\Delta t}{\pi}} \sum_{l=0}^{1}\left(\begin{array}{c}1\\l\end{array}\right)\sum_{i=0}^{1-l}\sum_{j=0}^{l} k^{(1-l,l)}_{i,j}e^{F_{i,j} (t-t_0)}\\
    \\
    (\Delta X_t)^2 - \Delta t \sum_{l=0}^{2}\left(\begin{array}{c}2\\l\end{array}\right)\sum_{i=0}^{2-l}\sum_{j=0}^{l} k^{(2-l,l)}_{i,j}e^{F_{i,j} (t-t_0)}\\
    \\  
    \abs{\Delta X_t}^3 - \sqrt{\frac{8\Delta t^3}{\pi}} \sum_{l=0}^{3}\left(\begin{array}{c}3\\l\end{array}\right)\sum_{i=0}^{3-l}\sum_{j=0}^{l} k^{(3-l,l)}_{i,j}e^{F_{i,j} (t-t_0)}\\
    \\
    \Delta X_t \Delta X_{t+\Delta t}^2  - \Delta t^2 \left(C_{2,0}+2C_{1,1}+C_{0,2}\right)^2 \mathcal{L}(L'\Delta t;t)\\
    \vdots\\
    \Delta X_t \Delta X_{t+L\Delta t}^2 - \Delta t^2 \left(C_{2,0}+2C_{1,1}+C_{0,2}\right)^2 \mathcal{L}(L''\Delta t;t)\\
    \\
    \Delta X_t^2 \Delta X_{t+K'\Delta t}^2 - \Delta t^2\times\mathrm{r.h.s.~of~Equation~(C.7)~for~}\tau=K'\Delta t\\
    \vdots\\
    \Delta X_t^2 \Delta X_{t+K''\Delta t}^2 - \Delta t^2\times\mathrm{r.h.s.~of~Equation~(C.7)~for~} \tau=K''\Delta t\\
    \\
  \end{array}
  }=\mathbf{0}\,,
$$
where $\Delta X_t=\ln{S_{t+\Delta t}}-\ln{{S_t}-\mu\Delta t}$, $\Delta
t=1/250~\mathrm{yr}$.
In the previous equation, the conditions referring to the return-volatility and
the squared return correlations depend on the four positive integers $L'<L''$,
$K'<K''$.
The choice of these values can be made based on a prior analysis of the time
scales of the correlation functions, and will be detailed later on.
Thus, the dimension $r$ of the vector
$\mathbf{h}(\boldsymbol{\theta},\boldsymbol{W}_{t})$ reduces to
$4+L''-L'+1+K''-K'+1$.
From an econometric point of view the problem of the estimation of parameters
is cast into a sound statistical framework.
By means of GMM we can obtain
an estimate of central values and associated statistical uncertainty for all
the unknowns of the problem.
However, the quantity to be optimized is highly non linear, the optimization
procedure of the twelve dimensional problem is \emph{per se} problematic, and finding
a solution under blind search can be extremely demanding.
For this reason, we prefer to proceed by invoking some reasonable arguments
concerning the nature of the problem under study.
The starting point of our heuristic is the observation that, until
now, we have devoted little attention to the role played by the parameter
$t_0$.
In principle it could be treated as an unknown parameter, however its role is quite different from that played by the others.
Since it mainly determines the regime of the factors processes, we assume
$t_0\rightarrow -\infty$ as done in the previous
work~\cite{PhysRevE.83.04111}.
Said differently, we assume that the data we are observing reflect stationary realizations
of $Y_t$ and $Z_t$.
Under this regime, mean-reverting processes do not depend on the initial time
values $y_0$ and $z_0$ any more, and we identify $y_\infty$ with $y_0$, and
$z_\infty$ with $z_0$.
Moreover, both $Y_t$ and $Z_t$ are unobserved processes reflecting the presence
in the market of investment strategies with heterogeneous time horizons.
Even though this assumption could be relaxed, it is plausible to assume that
the Brownian motions driving those processes are uncorrelated.
If we fix $\rYZ=0$ the problem greatly simplifies since all $F_{m,n}$ reduce
to $\FY_m+\FZ_n$, and all terms $C_{m,n}$ split into $C_{m,0}\times C_{0,n}$.
In~\cite{PhysRevE.83.04111} we prove that $\rXY<0$ suffices in order to reproduce the leverage effect. Since we do not want that the incorporation 
of the extra factor $Z_t$ has a relevant impact on the leverage, we fix $\rXZ$ equal to zero.
Finally, the considerations that follow \Eref{eq:XmomsExpand} in
Section~\ref{sec:themodel} have clarified the way the tail exponent of the
distribution of the volatility factors is responsible for the divergence of the
moments of $X_t$.
If $Y_t$ and $Z_t$ were characterized by two different tail exponents, the order of the first divergent moment of $X_t$
should be determined by the lowest of them.
In this respect the role played by
the highest exponent would be spoiled by the other one.
We therefore assume that the stationary distributions of $Y_t$ and $Z_t$ have
the same shape parameter $\nu=\nu_Y=\nu_Z$.
Now the reduced vector of
parameters reads
$\boldsymbol{\theta}^\mathrm{t}=\left(\mu,y_\infty,z_\infty,\tau_Y,\tau_Z,\rXY,\nu\right)$,
while the orthogonality relations simplify.
For instance, the first four relations reduce to 
\begin{eqnarray}
  &&\ang{\Delta X_t} = 0\,,\nonumber\\
  &&\ang{\abs{\Delta X_t} - \sqrt{\frac{2\Delta t}{\pi}}(y_\infty + z_\infty)}=0\,,\nonumber\\
  &&\ang{(\Delta X_t)^2   - (y_\infty + z_\infty)^2 \Delta t + \frac{y_\infty^2 + z_\infty^2}{\nu-2}\Delta t}=0 \,,\nonumber\\
  &&\ang{\abs{\Delta X}^3 - \sqrt{\frac{8\Delta t^3}{\pi}}\frac{(\nu-1)^2}{(\nu-3)(\nu-2)}(y_\infty^3 + z_\infty^3)}\nonumber\\
  &&\qquad-\ang{3\sqrt{\frac{8\Delta t^3}{\pi}}\frac{\nu-1}{\nu-2}(y_\infty+z_\infty)y_\infty z_\infty}=0\,, 
  \label{eq:orthogonal_mom}
\end{eqnarray}
and the numerator of the leverage for positive $\tau$ becomes
\begin{eqnarray*}
  &\rXY\sqrt{\frac{8}{\tau_Y(\nu-1)}}  \left\{\left[C^\mathrm{st}_{2,0}C^\mathrm{st}_{0,1}-\frac{\nu-1}{\nu-3}y_\infty C^\mathrm{st}_{1,0}C^\mathrm{st}_{0,1}\right]
  \exp{\left[-\left(1+\frac{\nu-2}{\nu-1}\right)\frac{\tau}{\tau_\mathcal{L}}\right]}\right.\nonumber\\
  & +\left[C^\mathrm{st}_{2,0}C^\mathrm{st}_{0,1}+C^\mathrm{st}_{1,0}C^\mathrm{st}_{0,2}
  -z_\infty\left(C^\mathrm{st}_{2,0}+C^\mathrm{st}_{1,0}C^\mathrm{st}_{0,1}\right)\right]
  \exp{\left[-\left(1+\frac{\tau_\mathcal{L}}{\tau_Z}\right)\frac{\tau}{\tau_\mathcal{L}}\right]}\nonumber\\ 
  & \left.+\left[\left(\frac{\nu-1}{\nu-3}y_\infty+z_\infty\right)\left(C^\mathrm{st}_{2,0}+C^\mathrm{st}_{1,0}C^\mathrm{st}_{0,1}\right)\right]
  \exp{\left(-\frac{\tau}{\tau_\mathcal{L}}\right)}\right\} \,,
  \label{eq:Lev-Stat}
\end{eqnarray*}
where the superscript $^{\mathrm{st}}$ stands for the stationary regime
corresponding to $t_0\rightarrow -\infty$, and we recall that
$\tau_\mathcal{L}=(\nu-1)\tau_Y/(\nu-2)$.
Even though the leverage correlation introduces a superposition of three
exponential functions, we have seen at the end of the
Section~\ref{sec:leverage} that the characteristic exponents are of the same
magnitude and are all dominated by $\tau_\mathcal{L}$.
For this reason, and recalling that the typical decay time for the leverage is
smaller than one hundred days, to perform the optimization we fix $L'=1$,
$L''=250$, and solve the problem for the first $254$ orthogonal relations.
This greatly enhances the convergence of the numerical algorithms.
Once an estimate of $\tau_\mathcal{L}$ is found, we fix $K'$ equal to two
times the integer part of $\tau_\mathcal{L}$, $K'=\lfloor
2\tau_\mathcal{L}\rfloor$~\footnote{In the following we adopt standard
mathematical notation $\lfloor \cdot \rfloor$ for the integer part function.}
and $K''=250$, and we perform the final optimization on the entire set of
$254+K''-K'+1$ orthogonal relations.
The latter relations, which correspond to the volatility structure, should be
fixed using the whole range for the lagged correlation, and not just the lags
above the leverage.
However, we preliminarily perform the numerical optimization with $K'$
running from one to the mentioned level and we see that the fit worsens.
The reason comes from the very structure of the expression which appears in the
r.h.s. of \Eref{eq:ACFnum3}.
As commented in Section~\ref{ssec:autocorr_vola}, the lagged correlation is
dominated by two time scales, the first one is of order of $\tau_\mathcal{L}$,
while the second one is the long run component.
When $K'=1$, we are fitting the behaviour of the whole curve, but for low lags
it is largely determined by the scaling of the leverage.
The results of our numerical explorations show that the low lags part of the
curve is hard to be reproduced while the long run component is always well
described.
When we reduce $K'$ to one the optimizer tries to catch up at short lags but at
the cost of an even worst distortion of the autocorrelation for intermediate
lag values, please refer to Figure~\ref{fig:ACorr} for a visual comparison of
the cases $K'=1$ and $K'=\lfloor 2\tau_\mathcal{L}\rfloor$.
In light of these results we decide to perform the GMM favouring the long run
behaviour.
An alternative and less abrupt approach could be a weighted optimization on the
entire curve with weights in the matrix $\hat{\boldsymbol{\Omega}}_T$ that
change smoothly from zero to one from low to high lag values.
Finally we perform the GMM just one time and we obtain
$\hat{\boldsymbol{\theta}}^{(1)}$ and $\hat{\boldsymbol{V}_T}/T$.
In order to compute a consistent estimate of $\sigma_Y^2$ and the associated
confidence level, we can extract a random sample from
$\mathrm{Normal}(\hat{\boldsymbol{\theta}}^{(1)},\hat{\boldsymbol{V}_T}/T)$ and
obtain a statistics of $\sigma_Y^2$ through the relation $2/(\tau_Y(\nu-1))$.
We proceed in an analogous way for $\sigma_Z^2$, and for
$\tau_\mathcal{L}=\tau_Y(\nu-1)/(\nu-2)$.
\begin{figure}[h]
  \caption{\label{fig:LCorr}Analytical description of the empirical leverage
correlation with values of parameters estimated by GMM with $K'=\lfloor
2\tau_\mathcal{L}\rfloor$.}
  \begin{center}
    \includegraphics[width=.79\textwidth]{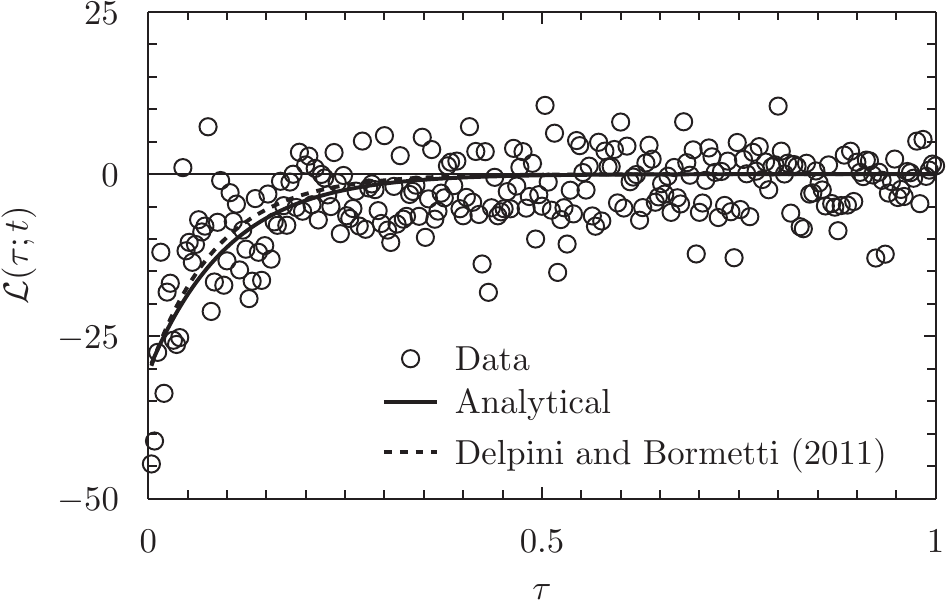}
  \end{center}
\end{figure}
The time series on which we perform the analysis is the same used
in~\cite{PhysRevE.83.04111}, and it consists of a data set from the Standard \&
Poor’s 500 index daily returns from 1970 to 2010.
This allows to evaluate the ability of the extended model to capture the
persistence of the volatility, not only in absolute terms but also in
comparison with the previous estimate from a simpler model.
\begin{table}
  \caption{\label{tab:parameters}Estimated values of the parameters
  ($K'=\lfloor 2\tau_\mathcal{L}\rfloor$) from daily returns of the S\&P500
index 1970-2010 .}
  \begin{center}
    \begin{tabular}{c|c|c|c}
      \rule[-7pt]{0pt}{19pt} & $\hat{\boldsymbol{\theta}}^{(1)}$ & $\hat{\boldsymbol{\sigma}}_T$ & $\hat{\boldsymbol{\rho}}_T$\\
      \hline
      &&&\\
      $\begin{array}{c}
	\mu        \\
	y_\infty   \\
	z_\infty   \\
	\tau_Y     \\
	\tau_Z     \\
	\rXY       \\
	\nu
      \end{array}$ 
      &
      $\begin{array}{l}
	2.1\times 10^{-4}\\
	0.095 \\
	0.052 \\
	0.07~\mathrm{yr}\\
	0.40~\mathrm{yr}\\
	-0.77  \\
	4.15
      \end{array}$
      & 
      $\begin{array}{l}
	6\times 10^{-5} \\
	0.004 \\
	0.004 \\
	0.01~\mathrm{yr}\\
	0.02~\mathrm{yr}\\
	0.09  \\
	0.01
      \end{array}$
      &
      $\left(\begin{array}{rrrrrrr}
	1.00 & -0.01 &  0.02 & -0.28 & -0.01 & -0.01 & -0.01 \\
	-0.01 &  1.00 & -0.97 & -0.04 & -0.14 &  0.00 &  0.99 \\
	0.02 & -0.97 &  1.00 &  0.03 &  0.25 &  0.00 & -0.94 \\
	-0.28 & -0.04 &  0.03 &  1.00 &  0.05 &  0.01 & -0.05 \\
	-0.01 & -0.14 &  0.25 &  0.05 &  1.00 &  0.00 & -0.12 \\
	-0.01 &  0.00 &  0.00 &  0.01 &  0.00 &  1.00 &  0.00 \\
	-0.01 &  0.99 & -0.94 & -0.05 & -0.12 &  0.00 &  1.00
      \end{array}\right)$\\
      &&&\\
      \hline
    \end{tabular}
  \end{center}
\end{table}
In \Tref{tab:parameters} we report the central values
$\hat{\boldsymbol{\theta}}^{(1)}$, the standard errors
$\hat{\boldsymbol{\sigma}}_T=\sqrt{\mathrm{diag}(\hat{\mathbf{V}}_T/T)}$, and the correlation structure
$\hat{\boldsymbol{\rho}}_T=\hat{\mathbf{V}}_T/(T\hat{\boldsymbol{\sigma}}_T\hat{\boldsymbol{\sigma}}_T^\mathrm{t})$
for all the parameters.
As far as the other relevant parameters of the model are concerned, we have
$\sigma_Y^2=9.9\pm 1.9$, $\sigma_Z^2=1.58\pm 0.07$, and
$\tau_\mathcal{L}=0.10\pm 0.02~\mathrm{yr}$.
The new values confirm the goodness of the estimate provided
in~\cite{PhysRevE.83.04111}, in particular the value of $\rXY$ is strictly
negative and the level of the tail exponent $\nu$ predicts the divergence of
moments higher than the fourth one.
More interesting to comment is the relationship between the different time
scales involved in our process.
Indeed, the shortest time scale corresponds to the typical relaxation time of
$Y_t$, which is found to be equal to $0.07~\pm~0.01~\mathrm{yr}$ and is
therefore dominated by the leverage time scale $0.10~\pm~0.02~\mathrm{yr}$ (to
be compared with the old estimate for $\tau_\mathcal{L}$
in~\cite{PhysRevE.83.04111} which is $0.09~\mathrm{yr}$).
The new time scale $\tau_Z$ for the process $Z_t$ is found to be a factor of
six larger than that of $Y_t$.
In Figures~\ref{fig:LCorr} and~\ref{fig:ACorr} we plot the leverage function
and the normalized autocorrelation of squared returns.
The exponential decay of the leverage is described correctly by the analytical
formula, and no relevant differences are noticeable with respect to the
description obtained via the model introduced in~\cite{PhysRevE.83.04111}.
Different considerations apply to the persistence of the volatility as
predicted by the extended model.
The presence of the slow volatility factor $Z_t$ introduces a longer time scale
allowing to capture the long range memory of the autocorrelation function.
This is evident from the comparison between the dashed line, corresponding to
the old model, and the bold one, corresponding to model~\eref{eq:model}.
Our results demonstrate the ability of a multi-factor approach to stochastic
volatility to effectively describe several phenomena.
In particular, even in the simplest version of a two factor model, it is able
to capture the emergence of multiple time scales for the volatility
autocorrelation as well as the exponential decay of the return-volatility
correlation.
The measured value for the tail parameter $\nu$ is coherent with the internal
consistency of the model requiring $\nu$ to be greater than four (in order for
\Eref{eq:ACFnum3} to converge in the stationary limit).
In particular, $\nu=4.15$ predicts an hyperbolic decay of the daily return
distribution which captures correctly the non Gaussian probability of extreme
events in the real data.

\section{Conclusions}\label{sec:conclusions}
In this work the model for the description of financial stylized facts proposed
in~\cite{PhysRevE.83.04111} is amended from the unrealistic fast decay of
the volatility autocorrelation.
This is achieved introducing an extra stochastic factor driving the
volatility.
In principle the number of factors could be increased at will, but the
analytical tractability of the resulting model would be hardly exploitable.
The intuition behind this generalization traces back to the early empirical
analysis of the FX market in~\cite{Muller_etal:1994} and the model
in~\cite{Muller_etal:1993}, where the role played by heterogeneous investors
is strongly emphasized.
Evidences from these papers are rooted in the econometric analysis of publicly 
available financial time series, but a convincing micro-founded model is still lacking.
Access to electronic order book data and to agents' identifiers would allow to
estimate the individual components of this heterogeneity.
An even approximate estimation of the distribution of typical investment
horizons from this information would provide a valuable trader-based
foundation.

With respect to previous approaches and analyses of continuous-time stochastic
volatility models, we believe that the calibration procedure proposed here
represents a further improvement and fulfils the desirable requirements of
statistical soundness.
At the same time, it also allows to focus on those facts which are 
established as relevant for the description of financial data.
In particular, we pursue an heuristic approach to optimization that,
reducing the dimensionality of the parameters space, retains only those
ingredients which are actually needed to capture the aforementioned empirical
evidences.

The stochastic volatility models discussed in this paper and
in~\cite{PhysRevE.83.04111} are inspired exclusively by the quest for a
realistic description of financial data.
In this quest, we focus on continuous time modelling and do not consider
aspects relating to possible applications in the financial sector.
In this respect, we should mention that important progresses have been obtained
by discrete time models.
In particular, models with multiscale ARCH volatility can also accommodate many
stylized facts, like the fat tails of returns, and reproduce consistently the
dynamics of realized volatility, also delivering accurate forecasts of the
latter~\cite{Zumbach:2011}.
They also prove to be very flexible tools for efficient option pricing and
hedging purposes~\cite{Zumbach:2012}.

In our case, the emergence of power-law tails in the return distribution
complies with past empirical analysis~\cite{Mantegna_book}, but also poses
serious limitations to the usage of the model in the context of option pricing.
This is certainly true for vanilla instruments, whose payoff grows
exponentially with the log-price.
On the other hand, our framework could deliver, in principle, better estimates
of the role played by rare events for market risk evaluation.

On the whole, we believe our model achieves a remarkable degree of realism,
higher than previous attempts in continuous-time stochastic volatility
modelling, yet allowing for important analytical derivations, e.g. that of the
moments of the return probability distribution.
Fairly enough, this comes at the price of elaborate manipulations and
Monte Carlo simulation would still be due for most financially relevant
applications.
Nonetheless recent advances in fast computing, e.g. GPU based numerical
techniques, could offer a promising scenario in this regard.

\ack
We thank the anonymous referees for constructive comments and suggestions.
We also warmly thank Fulvio Corsi for many inspiring discussions and we both 
acknowledge support of the Scuola Normale Superiore Grant `Giovani
Ricercatori 2011'.

\appendix
\section{Probability density function of price returns: moments computation}\label{app:A}
Application of 
It\^o's Lemma to the function $X_t^l$ readily provides
\begin{equation}
  \ang{X_t^l} =
  \frac{1}{2} l (l-1) \int_{0}^{t} \ang{X_s^{l-2} \left(Y_s+Z_s\right)^{2}}\,\ud s \,,
  \label{eq:Xmoments}
\end{equation}
and the same Lemma proves that the correlation functions between integer powers
of $X_t$, $Y_t$, and $Z_t$ satisfy the following differential equation
\begin{eqnarray}
  \frac{\ud}{\ud t} \ang{X_t^l Y_t^m Z_t^n} =
  F_{m,n} \,\ang{X_t^l Y_t^m Z_t^n} + \AY_m \,\ang{X_t^l Y_t^{m-1} Z_t^{n}}\nonumber\\
  + \AZ_n \,\ang{X_t^l Y_t^{m} Z_t^{n-1}}
  + \frac{1}{2} l (l-1) \,\ang{X_t^{l-2} Y_t^{m} Z_t^{n}\left(Y_t+Z_t\right)^2} \nonumber\\
  + \, l \left(m \, \rXY \sigma_Y + n \, \rXZ \sigma_Z \right)\, \ang{X_t^{l-1} Y_t^{m} Z_t^{n}\left(Y_t+Z_t\right)}\,,\nonumber\\
  \label{eq:corrDiffEq}
\end{eqnarray}
where the constants $F_{m, n}$ are defined right after
\Eref{eq:XmomsExpand} and $\AY_m=m\bY$ and $\AZ_n=n\bZ$.
Previous equations correspond to a system of nested linear ordinary
differential equation (ODE), which can be solved recursively starting from the
lowest order of $l$, $m$, and $n$, and whose solution involves integration of
the two point correlations $C_{m,n}(t;t_0) = \ang{Y_t^m Z_t^n}$~\footnote{In
the following we drop the dependence on $t$ and $t_0$.}.
From application of It\^o's Lemma we obtain
\begin{eqnarray*}
  \ud\left( Y^m \,Z^n \right) =
  \left[ \FY_m Y^m Z^n + \FZ_n Y^m Z^n \right] \,\ud t\\
  + \left[ \AY_m Y^{m-1} Z^n + \AZ_n Y^m Z^{n-1} \right] \,\ud t
  + \rYZ m n \sqrt{\cY \cZ} Y^m Z^n \,\ud t\\
  + m \sqrt{\cY} Y^m Z^n \,\ud W^Y_t + n \sqrt{\cZ} Y^m Z^n \ud W^Z_t \,;
\end{eqnarray*}
taking expectation, and differentiating \textit{w.r.t} time we derive the following ODE
\begin{eqnarray*}
  \frac{\ud}{\ud t} \ang{Y_t^m Z_t^n} =
  \left( \FY_m + \FZ_n + \rYZ m n \sqrt{\cY \cZ} \right) \ang{Y^m_t Z^n_t} \\
  + \AY_m \ang{Y_t^{m-1} Z_t^n} + \AZ_n \ang{Y_t^m Z_t^{n-1}} \,.
\end{eqnarray*}
For instance, for the case $m=n=1$ we have
\begin{eqnarray*}
  \frac{\ud}{\ud t} \ang{Y_t Z_t}= \left( \FY_1 + \FZ_1 + \rYZ \sqrt{\cY \cZ} \right)
  \ang{Y_t Z_t}\\
  + \AY_1 \ang{Z_t} + \AZ_1 \ang{Y_t}\,,
\end{eqnarray*}
where the mean values read 
\begin{eqnarray*}
  &\ang{Y_t} = -\frac{\AY_1}{\FY_1} + e^{\FY_1(t-t_0)}\left[ y_0 +\frac{\AY_1}{\FY_1} \right] \\
  &\ang{Z_t} = -\frac{\AZ_1}{\FZ_1} + e^{\FZ_1(t-t_0)}\left[ z_0 +\frac{\AZ_1}{\FZ_1} \right] \,,
\end{eqnarray*}
with $t_0\le 0$ the starting time of the factors processes.
More generally, by iterative solution it can be verified that $C_{m,n}$ admits
the expansion
\begin{equation}
  C_{m,n}=\ang{Y^m_t Z^n_t} = \sum_{i=0}^m \sum_{j=0}^n k^{(m,n)}_{i,j}e^{F_{i,j} (t-t_0)} \,,
  \label{eq:C2exp}
\end{equation}
where the coefficients depend on the initial conditions $y_0, z_0$ and satisfy
the recursive relations which follow
\begin{eqnarray}
  k^{(m,n)}_{i<m,j<n} & = & -\frac{\AY_m k^{(m-1,n)}_{i,j}+\AZ_n k^{(m,n-1)}_{i,j}}
  {F_{m,n}-F_{i,j}} \nonumber \\
  k^{(m,n)}_{i<m,n} & = & -\frac{\AY_m k^{(m-1,n)}_{i,n}}{F_{m,n}-F_{i,n}} \nonumber \\
  k^{(m,n)}_{m,j<n} & = & -\frac{\AZ_n k^{(m,n-1)}_{m,j}}{F_{m,n}-F_{m,j}} \nonumber \\
  k^{(m,n)}_{m,n} & = & \ang{Y^m_{t_0} Z^n_{t_0}}
  +\AY_m \sum_{i=0}^{m-1} \sum_{j=0}^n \frac{k^{(m-1,n)}_{i,j}}{F_{m,n}-F_{i,j}} \nonumber \\
  &&+\AZ_n \sum_{i=0}^{m} \sum_{j=0}^{n-1} \frac{k^{(m,n-1)}_{i,j}}{F_{m,n}-F_{i,j}} \,.
  \label{eq:kCoeffs}
\end{eqnarray}
We notice that the moments $\mu^Y_m(t)=\ang{Y_t^m}$ and
$\mu^Z_n(t)=\ang{Z_t^n}$ are specific cases of the
expansion~\eref{eq:C2exp}, whose coefficients are given by the column vector
$(k^{(m,0)}_{i,0})_{i\le m}$ and the row vector $(k^{(0,n)}_{0,j})_{j\le n}$,
while in general the set of coefficients $k^{(m,n)}_{i,j}$ can be cast
in a $(m+1) \times (n+1)$ real matrix.
For instance, for the case $C_{2,1}=\ang{Y^2_t Z_t}$ we obtain
\begin{eqnarray*}
  k^{(2,1)}_{0,0} & = & -\frac{1}{F_{2,1}} \left[ \AY_2 k^{(1,1)}_{0,0}+\AZ_1 k^{(2,0)}_{0,0} \right] \,,\\
  k^{(2,1)}_{0,1} & = & -\frac{\AY_2 k^{(1,1)}_{0,1}}{F_{2,1}-\FZ_1} \,,\\
  k^{(2,1)}_{1,0} & = & -\frac{1}{F_{2,1}-\FY_1} \left[ \AY_2 k^{(1,1)}_{1,0}+\AZ_1 k^{(2,0)}_{1,0} \right] \,,\\
  k^{(2,1)}_{1,1} & = & -\frac{\AY_2 k^{(1,1)}_{1,1}}{F_{2,1}-F_{1,1}} \,,\\
  k^{(2,1)}_{2,0} & = & -\frac{\AZ_1 k^{(2,0)}_{2,0}}{F_{2,1}-F_{2}} \,,\\
  k^{(2,1)}_{2,1} & = & \ang{Y_{t_0}^2 Z_{t_0}} - \left[ k^{(2,1)}_{0,0} + k^{(2,1)}_{0,1} + k^{(2,1)}_{1,0} + k^{(2,1)}_{1,1} + k^{(2,1)}_{2,0}\right] \,.
\end{eqnarray*}
Ultimately, given the expansion \eref{eq:C2exp} and by inspection of
\eref{eq:Xmoments}, we recognize that the moments of $X$ can be cast in the
form of \Eref{eq:XmomsExpand}.

\section{Computation of the return-volatility correlation}\label{app:B}
The results which follow are derived under the assumption that $\rXZ$ is equal to zero.
The numerator of the return-volatility correlation $\mathcal{L}(\tau;t) =
\ang{\ud X_t \,\ud X_{t+\tau}^2}/\ang{\ud X_t^2}^2$ can be cast in the form 
\begin{equation*}
  \ang{\ud X_t \,\ud X_{t+\tau}^2} = \ang{\left( Y_t + Z_t \right)\left( Y_{t+\tau} + Z_{t+\tau} \right)^2\zeta_t^X}\,\ud t^2 \,.
\end{equation*}
Here, adopting the same convention of~\cite{Perello:2004}, we
formally~\footnote{This convention is somewhat unusual, but admissible as the
distributional assumption on $\zeta^X_t$ guarantees that $dW_t$ is still
Gaussian with zero mean and variance equal to $\ud t$.
This choice proves to be
convenient to develop the following calculations.} express the Wiener
increment as $\ud W_t^X=\zeta_t^X \,\ud t$, where $\zeta_t^X$ is a Gaussian
noise with zero mean and variance $\ang{ (\zeta_t^X)^2 } = 1 / \ud t$.
Novikov's theorem~\cite{SovPhys.20.1290,PhysRevE.67.037102} allows to
compute the expectation involving $\zeta_t^X$, giving us
\begin{eqnarray}
  &\frac{\ang{\ud X_t \,\ud X_{t+\tau}^2}}{\ud t^2} =
  ~ 2 \rXY\sqrt{\cY} H(\tau)\exp{\left(\aY \tau\right)}\times \nonumber\\ 
  &\qquad\ang{\left[Y_t(Y_t + Z_t) (Y_{t+\tau} + Z_{t+\tau}) \right]
  \exp{\left[\sqrt{\cY} \Delta_t W^Y(\tau)\right]}}\,,
  \label{eq:Lev-Num}
\end{eqnarray}
where we define $ \Delta_t W(\tau) \doteq \int_{t}^{t+\tau}\,dW_s$.
We refer the interested reader to Section IV in~\cite{PhysRevE.83.04111} for
further details regarding the derivation of the previous equation. The right
hand side of~\eref{eq:Lev-Num} can be split into four pieces proportional to
the expectations 
\begin{eqnarray*}
	f_{YYY}(\tau,t)& \doteq & \ang{Y_t^2 \,Y_{t+\tau} \exp{\left[\sqrt{\cY}\,\Delta_t W^Y(\tau)\right]}}\,,\\
	f_{YYZ}(\tau,t)& \doteq & \ang{Y_t^2 \,Z_{t+\tau} \exp{\left[\sqrt{\cY}\,\Delta_t W^Y(\tau)\right]}}\,,\\
	f_{YZY}(\tau,t)& \doteq & \ang{Y_t \,Z_t \,Y_{t+\tau} \exp{\left[\sqrt{\cY}\,\Delta_t W^Y(\tau)\right]}}\,,\\
	f_{YZZ}(\tau,t)& \doteq & \ang{Y_t \,Z_t \,Z_{t+\tau} \exp{\left[\sqrt{\cY}\,\Delta_t W^Y(\tau)\right]}}\,.
\end{eqnarray*}
Following the approach discussed in Appendix B of ~\cite{PhysRevE.83.04111}, it
is possible to show that they satisfy the relations
\begin{eqnarray*}
	f_{YYY}(\tau,t) -(\cY\aY) \int_0^\tau f_{YYY}(\tau',t)\exp{\left[\frac{\cY}{2}(\tau-\tau')\right]}\ud\tau' = \nonumber\\ 
	\qquad\exp{\left(\frac{\cY}{2}\tau\right)}\left[ C_{3,0}+\bY\tau C_{2,0} \right]\,,\nonumber\\
	f_{YYZ}(\tau,t) +\kappa_Z \int_0^\tau f_{YYZ}(\tau',t)\exp{\left[\frac{\cY}{2}(\tau-\tau')\right]}\ud\tau' = \nonumber\\ 
	 \qquad\exp{\left(\frac{\cY}{2}\tau\right)}\left[ C_{2,0} C_{0,1}+\bZ\tau C_{2,0} \right]\,,\nonumber\\
	f_{YZY}(\tau,t) -(\cY\aY) \int_0^\tau f_{YZY}(\tau',t)\exp{\left[\frac{\cY}{2}(\tau-\tau')\right]}\ud\tau' = \nonumber\\ 
	 \qquad\exp{\left(\frac{\cY}{2}\tau\right)}\left[ C_{2,0} C_{0,1}+\bY\tau C_{1,0} C_{0,1} \right]\,,\nonumber\\
	f_{YZZ}(\tau,t) +\kappa_Z \int_0^\tau f_{YZZ}(\tau',t)\exp{\left[\frac{\cY}{2}(\tau-\tau')\right]}\ud\tau' = \nonumber\\ 
	 \qquad\exp{\left(\frac{\cY}{2}\tau\right)}\left[ C_{1,0} C_{0,2}+\bZ\tau C_{1,0} C_{0,1} \right]\,,\nonumber
\end{eqnarray*}
corresponding to a set of Volterra integro-differential equations of the second
kind.
Their solutions are known in closed-form, and after plugging them in
\Eref{eq:Lev-Num}, the final expression of the leverage correlation reads
\begin{eqnarray}
    \mathcal{L}(\tau;t) = \frac{2 \rXY\sqrt{\cY} H(\tau)}{\left(C_{2,0}+2C_{1,0}C_{0,1}+C_{0,2}\right)^2}\times\nonumber\\
    \left\{\left[C_{3,0}+C_{2,0}C_{0,1}+\frac{\bY}{\cY\aY}(C_{2,0}+C_{1,0}C_{0,1})\right]\exp{\left[2\left(\frac{3\cY}{4}\aY\right)\tau\right]}\right.\nonumber\\
    +\left[C_{2,0}C_{0,1}+C_{1,2}-z_\infty\left(C_{2,0}+C_{1,0}C_{0,1}\right)\right]\exp{\left[\left(\frac{\cY}{2}\aY\aZ\right)\tau\right]}\nonumber\\ 
    \left.-\left[\left(\frac{\bY}{\cY\aY}-z_\infty\right)\left(C_{2,0}+C_{1,0}C_{0,1}\right)\right]\exp{\left[\left(\frac{\cY}{2}\aY\right)\tau\right]}\right\} \,.
    \label{eq:Lev}
\end{eqnarray}
A meaningful comparison of the previous expression with real data requires to
take its stationary limit for $t_0\to -\infty$, which amounts to replacing
$C_{m,0} C_{0,n}$ with the asymptotic values $C_{m,0}^\mathrm{st} C_{0,n}^\mathrm{st}$.

\section{Computation of the squared return correlation}\label{app:C}
Resorting to the same parametrization of the Wiener variation adopted in
\ref{app:B} 
we have
\begin{eqnarray}
  \ang{\ud X_t^2 \,\ud X_{t+\tau}^2} & = & \ud t^2 \ang{ \left( Y_t + Z_t \right)^2 \left( Y_{t+\tau} + Z_{t+\tau} \right)^2
  \,\ud W^X_t \,\zeta^X_t } \nonumber \\
  & = & \ud t^2 \ang{ \left( Y_t + Z_t \right)^2 \left( Y_{t+\tau} + Z_{t+\tau} \right)^2 } + \mathcal{O}(\ud t^3) \,.
  \label{eq:ACFnum}
\end{eqnarray}
In order to compute the autocorrelation function of squared returns,
the quantities $f^{(m,n,p,q)}_t(\tau)=\ang{Y_t^m Z_t^n Y^{p}_{t+\tau}
Z^{q}_{t+\tau}}$ indicating the $\tau$-lagged correlation have to be evaluated.
The relevant cases correspond to $p,q\le 2$, and below we detail the
corresponding exact results, all of which are obtained replacing the process
$Y^p_{t+\tau} Z^q_{t+\tau}$ with its integral representation from time $t$ to
time $t+\tau$.
\paragraph{Computation of $f^{(m,n,1,0)}_t(\tau)=\ang{Y^m_t Z^n_t Y_{t+\tau}}$.}
It is readily verified that $f^{(m,n,1,0)}_t(\tau)$ is solution of a linear ODE, giving
\begin{equation}
  f^{(m,n,1,0)}_t(\tau) = -\frac{\AY_1}{\FY_1} C_{m,n}
  + e^{\FY_1 \tau} \left[ C_{m+1,n}+\frac{\AY_1}{\FY_1} C_{m,n} \right]\,.
  \label{eq:f10}
\end{equation}
\paragraph{Computation of $f^{(m,n,0,1)}_t(\tau)=\ang{Y^m_t Z^n_t Z_{t+\tau}}$.}
In much the same way we have
\begin{equation*}
  f^{(m,n,0,1)}_t(\tau) = -\frac{\AZ_1}{\FZ_1} C_{m,n}
  + e^{\FZ_1 \tau} \left[ C_{m,n+1}+\frac{\AZ_1}{\FZ_1} C_{m,n} \right]\,.
\end{equation*}
\paragraph{Computation of $f^{(m,n,2,0)}_t(\tau)=\ang{Y^m_t Z^n_t Y^2_{t+\tau}}$.}
After replacement of $Y^2_{t+\tau}$, we can write
\begin{eqnarray*}
  f^{(m,n,2,0)}_t(\tau) &=& f^{(m,n,2,0)}_t(0)
  + \FY_2 \int_0^{\tau} f^{(m,n,2,0)}_t(\tau') \ud\tau' \\
  &&+\AY_2 \int_0^{\tau} f^{(m,n,1,0)}_t(\tau') \ud\tau' \,;
\end{eqnarray*}
further, we can replace the solution~\eref{eq:f10} for $f^{(m,n,1,0)}_t(\tau')$
in the second integral, leading to straightforward integrations of exponential
functions of $\tau$.
Finally, we are left with
\begin{eqnarray}
  &f^{(m,n,2,0)}_t(\tau)  =  \left[ \frac{\AY_2 \AY_1}{\FY_2 \FY_1} C_{m,n} \right]\nonumber\\
  &\qquad + e^{\FY_1 \tau} \left[ -\frac{\AY_2}{\FY_2-\FY_1} \left( C_{m+1,n}
  +\frac{\AY_1}{\FY_1} C_{m,n} \right) \right] \nonumber \\
  &\qquad +e^{\FY_2 \tau} \left[ C_{m+2,n}
  +\frac{\AY_2}{\FY_2-\FY_1} \left( C_{m+1,n}
  +\frac{\AY_1}{\FY_2} C_{m,n} \right) \right]\,.
  \label{eq:f20}
\end{eqnarray}
\paragraph{Computation of $f^{(m,n,0,2)}_t(\tau)=\ang{Y^m_t Z^n_t Z^2_{t+\tau}}$.}
As before, after replacement of the parameters for the dynamics of the
$Z_{t+\tau}^2$ process, we get to
\begin{eqnarray}
  &f^{(m,n,0,2)}_t(\tau) = \left[ \frac{\AZ_2 \AZ_1}{\FZ_2 \FZ_1} C_{m,n} \right] \nonumber\\
  &\qquad + e^{\FZ_1 \tau} \left[ -\frac{\AZ_2}{\FZ_2-\FZ_1} \left( C_{m,n+1}
  +\frac{\AZ_1}{\FZ_1} C_{m,n} \right) \right] \nonumber \\
  &\qquad + e^{\FZ_2 \tau} \left[ C_{m,n+2}
  +\frac{\AZ_2}{\FZ_2-\FZ_1} \left( C_{m,n+1}
  +\frac{\AZ_1}{\FZ_2} C_{m,n} \right) \right]\,.
  \label{eq:f02}
\end{eqnarray}
\paragraph{Computation of $f^{(m,n,1,1)}_t(\tau)=\ang{Y^m_t Z^n_t Y_{t+\tau} Z_{t+\tau}}$.}
The evolution of the joint process $Y_t Z_t$ is given by
\begin{eqnarray*}
  \ud (Y_t Z_t) & = & \left( F_{1,1} Y_t Z_t + \AZ_1 Y_t + \AY_1 Z_t \right) \,\ud t \\
  &&+\sqrt{\cY} Y_t Z_t \,\ud W^Y_t + \sqrt{\cZ} Y_t Z_t \ud W^Z_t\,,
\end{eqnarray*}
and substitution inside the expectation gives
\begin{eqnarray}
  &f^{(m,n,1,1)}_t(\tau) = \left[ \frac{\AY_1 \AZ_1}{F_{1,1}}
  \left( \frac{1}{\FY_1}+\frac{1}{\FZ_1} \right)\right] C_{m,n} \nonumber \\
  &\quad- \frac{\AZ_1}{F_{1,1}-\FY_1} \,e^{\FY_1 \tau}
  \left[ C_{m+1,n}+\frac{\AY_1}{\FY_1} C_{m,n} \right] \nonumber\\
  &\quad- \frac{\AY_1}{F_{1,1}-\FZ_1} \,e^{\FZ_1 \tau}
  \left[ C_{m,n+1}+\frac{\AZ_1}{\FZ_1} C_{m,n} \right]  \nonumber \\
  &\quad+ e^{F_{1,1} \tau}
  \left[ C_{m+1,n+1} + \frac{\AZ_1}{F_{1,1}-\FY_1} C_{m+1,n}
  +\frac{\AY_1}{F_{1,1}-\FZ_1} C_{m,n+1} \right.  \nonumber \\
  &\quad- \left. \AY_1 \AZ_1 \left( \frac{2F_{1,1}-\FY_1-\FZ_1}{F_{1,1}(F_{1,1}-\FY_1)(F_{1,1}-\FZ_1)} \right) C_{m,n} \right] \,.
  \label{eq:f11}
\end{eqnarray}
As expected from the structure of model~\eref{eq:model}, and as confirmed by all previous examples, it is clear that the
functions $f^{(m,n,p,q)}_t(\tau)$ admit a general expansion reading 
\begin{equation*}
  f^{(m,n,p,q)}_{t}(\tau) = \sum_{i=1}^{p} \sum_{j=1}^q
  h^{(m,n,p,q)}_{i,j}(t) \,e^{F_{i,j} \tau}\,,
\end{equation*}
where the terms $h^{(m,n,p,q)}_{i,j}(t)$ can be computed exactly.
Coming back to \Eref{eq:ACFnum} we have
\begin{eqnarray}
    \frac{\ang{\ud X_t^2 \,\ud X_{t+\tau}^2}}{\ud t^2} = \nonumber\\
    ~ f_t^{(2,0,2,0)}(\tau) + f_t^{(0,2,2,0)}(\tau) + 2f_t^{(0,2,2,0)}(\tau) + f_t^{(2,0,0,2)}(\tau) + f_t^{(0,2,0,2)}(\tau) \nonumber \\
    + 2f_t^{(0,2,0,2)}(\tau) + 2\left[ f_t^{(2,0,1,1)}(\tau) + f_t^{(0,2,1,1)}(\tau) + 2f_t^{(0,2,1,1)}(\tau) \right] \,.
    \label{eq:ACFnum2}
\end{eqnarray}
By means of Equations~\eref{eq:f20}-\eref{eq:f11}, and after defining the
auxiliary variables
\begin{eqnarray*}
  &T_1 = C_{2,0}+C_{0,2}+2C_{1,1}\,,\quad T_2 = C_{3,0}+C_{1,2}+2C_{2,1}\,,\\
  &T_2^* = C_{0,3}+C_{2,1}+2C_{1,2}\,,\quad T_3 = C_{4,0}+C_{2,2}+2C_{3,1}\,,\\
  &T_3^* = C_{0,4}+C_{2,2}+2C_{1,3}\,,\quad T_4 = C_{3,1}+C_{1,3}+2C_{2,2}\,,
\end{eqnarray*}
we can write the following final expression
\begin{eqnarray}
  \frac{\ang{\ud X_t^2 \,\ud X_{t+\tau}^2}}{\ud t^2} =&
  \left[ \frac{\AY_2 \AY_1}{\FY_2 \FY_1} + \frac{\AZ_2 \AZ_1}{\FZ_2 \FZ_1} + 2 \frac{\AY_1 \AZ_1}{F_{1,1}}\left( \frac{1}{\FY_1}+\frac{1}{\FZ_1} \right) \right] T_1 \nonumber\\
  &-e^{\FY_1 \tau} \left( T_2 + \frac{\AY_1}{\FY_1} T_1 \right) \left[ \frac{\AY_2}{\FY_2-\FY_1}+2 \frac{\AZ_1}{F_{1,1}-\FY_1} \right] \nonumber\\
  & -e^{\FZ_1 \tau} \left( T_2^* + \frac{\AZ_1}{\FZ_1} T_1 \right) \left[ \frac{\AZ_2}{\FZ_2-\FZ_1}+2 \frac{\AY_1}{F_{1,1}-\FZ_1} \right] \nonumber\\
   &+e^{\FY_2 \tau} \left[ T_3 + \frac{\AY_2}{\FY_2-\FY_1} \left( T_2 +\frac{\AY_1}{\FY_2} T_1 \right) \right]\nonumber \\
   &+e^{\FZ_2 \tau} \left[ T_3^* + \frac{\AZ_2}{\FZ_2-\FZ_1} \left( T_2^* +\frac{\AZ_1}{\FZ_2} T_1 \right) \right] \nonumber\\
   &+2 e^{F_{1,1} \tau} \left[ T_4 + \frac{\AY_1}{F_{1,1}-\FZ_1} T_2^* + \frac{\AZ_1}{F_{1,1}-\FY_1} T_2 \right. \nonumber\\
   &\left. +\frac{2F_{1,1}-\FY_1-\FZ_1}{F_{1,1}(F_{1,1}-\FY_1)(F_{1,1}-\FZ_1)} \right] \,.
  \label{eq:ACFnum3}
\end{eqnarray}
Ultimately, evaluation of the volatility autocorrelation~\eref{eq:AcorrDef}
requires to compute $\mathrm{Var}[\ud X_t^2] = \ang{\ud X_t^4}-\ang{\ud
X_t^2}^2$ which is given by 
\begin{equation*}
  3\left( C_{4,0}+4C_{3,1}+6C_{2,2}+4C_{1,3}+C_{0,4} \right) \ud t^2
  -\left( C_{2,0}+2C_{1,1}+C_{0,2} \right)^2\ud t^2\,.
\end{equation*}

\section*{References}
\providecommand{\newblock}{}

\end{document}